\documentclass[prb,aps]{revtex4}
\usepackage{graphicx}
\begin{document}

\title{Criticality and heterogeneity in the solution space of random constraint satisfaction problems}

\author{Haijun Zhou\footnote{Paper accepted for publication by {\it International Journal of Modern Physics B}}}

\affiliation{Key Laboratory of Frontiers in Theoretical Physics and
	Kavli Institute for Theoretical Physics China, 
	Institute of Theoretical Physics, Chinese Academy of Sciences,
		Beijing 100190,	China}

\date{\today}

\begin{abstract}
Random constraint satisfaction problems are  interesting model systems
for spin-glasses and glassy dynamics studies.  
As the constraint density of such a system reaches  certain
threshold value, its solution space may split into extremely many 
clusters.
In this work we argue that this ergodicity-breaking
transition is preceded by a homogeneity-breaking transition.
For random $K$-SAT and $K$-XORSAT, we
show that many solution communities start to form in the solution space
as the constraint density reaches a critical value $\alpha_{cm}$, with
each community containing a set of solutions
that are more similar with each other than with the outsider solutions.
At $\alpha_{cm}$ the solution space is in a critical state. 
The connection of these results to the onset of dynamical heterogeneity in
lattice glass models is discussed. 
\end{abstract}


\maketitle

\section{Introduction}

Random constraint satisfaction problems (CSPs) 
have attracted a lot of research interest from the statistical physics community
in recent years\cite{Hartmann-Weigt-2005,Mezard-Montanari-2009,Monasson-Zecchina-1996,Mezard-etal-2002}.
These  are  model systems for understanding typical-case
computational complexity of nonpolynomial-complete problems in computer science,
some of them have also important applications in modern coding systems,
such as the low-density-parity-check codes. The energy landscape
of a nontrivial random CSP problem is usually very complicated, similar to those of
spin-glasses\cite{Mezard-Parisi-2001,Mezard-Parisi-2003}
and lattice glass models\cite{Biroli-Mezard-2002,Darst-Reichman-Biroli-2009}.
Therefore understanding the configuration space property of random CSPs is also
very helpful for developing new insights for spin-glasses, glassy dynamics, and
the jamming phenomena of colloids and granular systems.

As the density $\alpha$ of constraints increases, the solution space of a CSP
will experience a series of phase transitions\cite{Krzakala-etal-PNAS-2007}. One of them
is the clustering transition at certain threshold
constraint density $\alpha_d$,  where the solution space splits
into exponentially many isolated solution clusters or Gibbs states.
This ergodicity-breaking transition has very significant consequences for
glassy dynamics\cite{Montanari-Semerjian-2006,Mezard-Montanari-2006} and 
stochastic local search processes\cite{Krzakala-Kurchan-2007,Alava-etal-2008,Zhou-2009}.
Numerical simulations\cite{Zhou-Ma-2009} further
suggested that, before ergodicity of the
solution space is broken, the solution space has already been non-homogeneous,
with the
formation of many solution communities. In this paper we determine the critical
constraint density $\alpha_{cm}$ for the solution space of a random CSP
to become heterogeneous. We find that $\alpha_{cm}< \alpha_d$, and that at $\alpha=\alpha_{cm}$
the solution space is in a critical state, in which the boundaries between different
solution communities of the solution space disappears, while at $\alpha> \alpha_{cm}$ the
solution space contains many well-formed solution communities.

Heterogeneity of the configuration space of a complex system can cause
heterogeneity in the dynamics of this system. The results of this work
may be helpful for understanding more quantitatively the nature of
dynamical heterogeneity in supercooled liquids\cite{Ediger-2000,Glotzer-2000,Cavagna-2009} 
and lattice glass models\cite{Darst-Reichman-Biroli-2009,Biroli-Mezard-2002}.


\section{Theory}

A constraint satisfaction formula has $N$ vertices
($i,j,k,\ldots$) and $M$ constraints ($a,b,c,\ldots$),
with constraint density  $\alpha \equiv M/N$.
A configuration of the model is denoted by
$\vec{\sigma} \equiv \{ \sigma_1, \sigma_2, \ldots, \sigma_N\}$,
where $\sigma_i = \pm 1$ is the spin state of vertex $i$.
Each constraint $a$ represents a multi-spin interaction among
a subset (denoted as $\partial a$) of vertices, and its 
energy $E_a$ is either zero (constraint being satisfied) or
positive (unsatisfied). For example, $E_a$ may be expressed
as
\begin{equation}
	\label{eq:xorsat}
	E_a = 1- J_a \prod\limits_{i \in \partial a} \sigma_i
\end{equation}
where $J_a = +1$ or $-1$ depending on the constraint $a$.
The total energy of a
spin configuration $\vec{\sigma}$ is the sum of individual
constraint energies,
$E(\vec{\sigma}) = \sum_{a=1}^{M} E_a$.

A solution of a constraint satisfaction formula is a spin configuration
of zero total energy. The whole set of solutions for a
given energy function $E(\vec{\sigma})$ is denoted as
$\mathcal{S}$ and referred to as the solution space.
The energy landscape of a CSP may be very complex and
its ground-state degeneracy usually is extremely high.
To investigate the
structure of a solution space $\mathcal{S}$, 
we define a partition function  as
\begin{equation}
	\label{eq:Partition_function}
	Z(x) = \sum\limits_{\vec{\sigma}^{1} \in \mathcal{S}}
	\sum\limits_{\vec{\sigma}^{2} \in \mathcal{S}}
	\exp\Bigl( x \sum\limits_{i=1}^{N} \sigma_i^{1}
	 \sigma_i^{2} \Bigr)
	\ ,
\end{equation}
where $x$ is a binding field. Each solution pair
$(\vec{\sigma}^{1}, \vec{\sigma}^{2})$ contributes a term
$\exp[N x q(\vec{\sigma}^{1}, \vec{\sigma}^{2})]$ to 
$Z(x)$, with $q(\vec{\sigma}^{1}, \vec{\sigma}^{2})
\equiv (1/N) \sum_{i=1}^{N} \sigma_i^{1} \sigma_i^{2}$ being the
solution-solution overlap.
We introduce an entropy density $s(q)$ by expressing
the total number $\mathcal{N}(q)$ of solution-pairs with
overlap $q$ as $\mathcal{N}(q)=\exp[N s(q)]$. Then
Eq.~(\ref{eq:Partition_function}) is re-written as
\begin{equation}
	\label{eq:partition_function_2}
	Z(x)= \sum\limits_{q} \exp\Bigl[ N \bigl(
	s(q) + x q \bigr) \Bigr] \ .
\end{equation}
The free entropy\cite{Mezard-Montanari-2009}
$\Phi(x)$ of the system is related to the partition
function by $\Phi(x)\equiv \ln Z(x)$
In the thermodynamic limit of $N\rightarrow \infty$,
the free entropy density $\phi(x)\equiv \Phi(x)/N$ is related to
the entropy density $s(q)$ by
\begin{equation}
	\label{eq:free_entropy_density}
	\phi(x)= \max\limits_{q\in[-1,1]}
	\Bigl[s(q)+ x q \Bigr] = s\bigl(\overline{q}(x)\bigr)
	+x \overline{q}(x) \ .
\end{equation}
In Eq.~(\ref{eq:free_entropy_density}), $\overline{q}(x)$ is the
overlap value at which the function $s(q)+x q$ 
achieves the global maximal value.
$\overline{q}(x)$ is also the mean value of solution-solution overlaps 
under the binding field $x$. 

%
\begin{figure}[t]
	\begin{center}
		\includegraphics[width=0.5\textwidth]{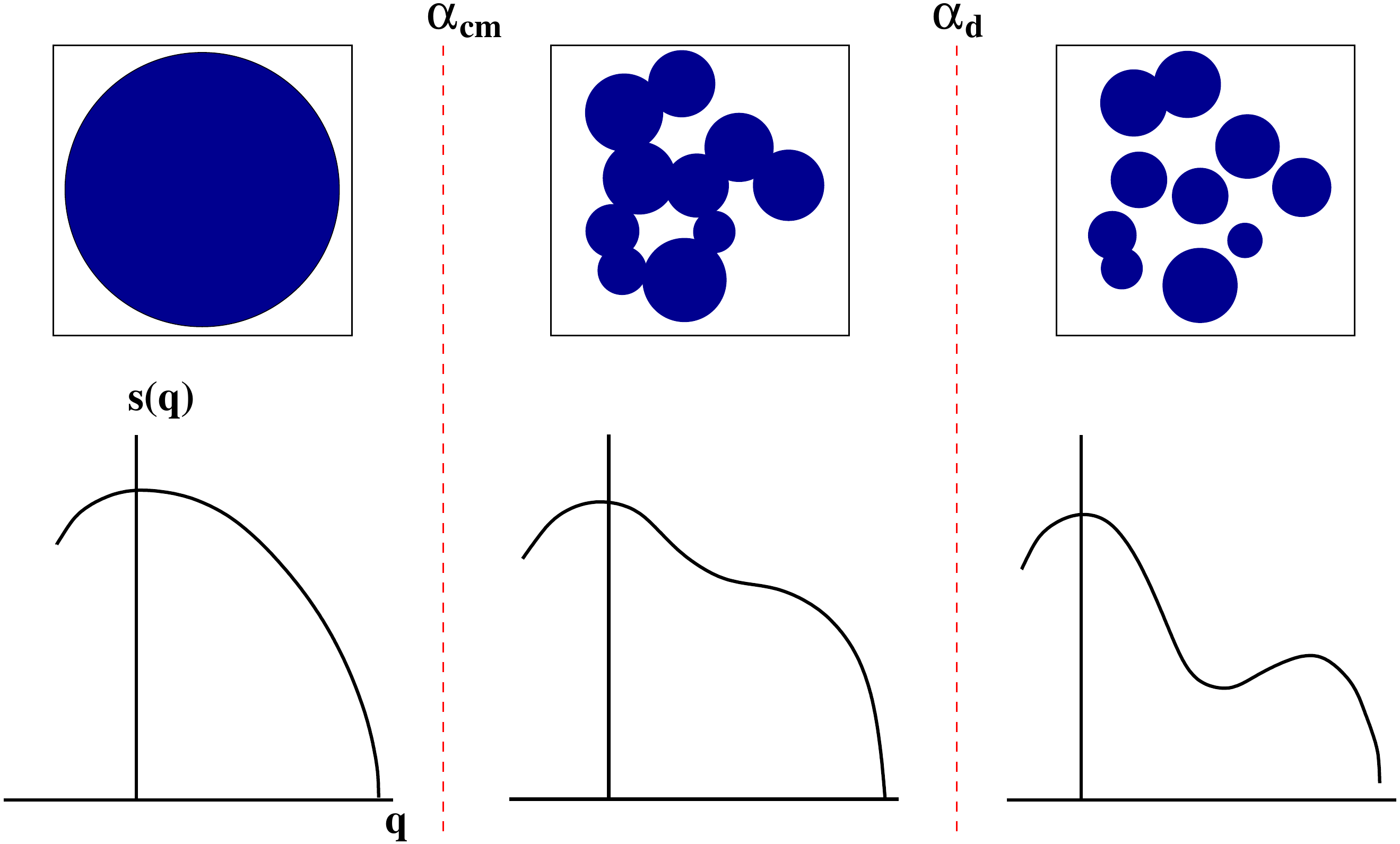}
	\end{center}
	\caption{\label{fig:community}
	Evolution of the solution space $\mathcal{S}$
	of a constrained spin system:
	At low constraint density $\alpha$ (left panel), $\mathcal{S}$ is
	homogeneous and the solution-pair entropy density $s(q)$ is
	a concave function of the overlap $q$. Solution communities start
	to form as $\alpha$ exceeds a threshold value $\alpha_{cm}$ 
	(middle panel); $\mathcal{S}$ then
	becomes heterogeneous and  the function
	$s(q)$ changes to be non-concave. An ergodicity-breaking
	transition occurs as $\alpha$ reaches a larger threshold value
	$\alpha_{d}$ (right panel), where the solution communities
	separate into  different solution clusters and $s(q)$ becomes
	non-monotonic. As $\alpha$ further increases, solution-pairs with
	intermediate overlap values may disappear completely, and then $s(q)$
	is not defined for these intermediate overlap values.
	}
\end{figure}

At $x=0$, the maximum of Eq.~(\ref{eq:free_entropy_density})
is achieved at $\overline{q}(0)=q_0$, the most probable 
solution-pair overlap value; at the other limit of $x\rightarrow \infty$,
$\overline{q}(\infty)=1$.
If the entropy density $s(q)$ is a concave function of $q \in [q_0, 1]$
(Fig.~\ref{fig:community}, left panel), then for
each  $x>0$ there is only one mean overlap
value $\overline{q}$, and $\overline{q}(x)$ changes smoothly with $x$. 
On the other hand, if $s(q)$ is non-concave
in $q\in [q_0, 1]$ (Fig.~\ref{fig:community}, middle and right panel),
then at certain value $x^*$ of
the binding field, there are two different mean overlap values, and
the value of $\overline{q}(x)$ changes {\em discontinuously} at $x=x^*$
(a field-induced first-order phase-transition).
In this work, we exploit
this correspondence between the non-concavity of $s(q)$ and the
discontinuity of $\overline{q}(x)$ to determine the
threshold constraint density $\alpha_{cm}$ at which
the solution space $\mathcal{S}$ becomes heterogeneous. Many solution
communities can be identified in a heterogeneous solution space
$\mathcal{S}$\cite{Zhou-Ma-2009}. Each solution community contains a
set of solutions which are more similar with each other than
with the solutions of other communities. These differences of
intra- and inter-community overlap values and the relative sparseness
of solutions at the boundaries between solution communities
cause the non-concavity
of $s(q)$.


\section{Application to the random $K$-SAT problem}

We begin with the random $K$-SAT, a
prototypical CSP\cite{Monasson-Zecchina-1996,Mezard-etal-2002,Krzakala-etal-PNAS-2007}.
In a random $K$-SAT formula, the number of vertices in
the set $\partial a$ of each constraint $a$ is fixed to $K$,
and these $K$ different vertices are randomly chosen from the
whole set of $N$ vertices. Depending on the spins of these $K$ vertices,
the energy of a constraint $a$ is either zero or unity:
\begin{equation}
	\label{eq:energy-sat}
	E_a=\prod\limits_{i\in \partial a} \frac{1-J_a^i \sigma_i}{2} \ ,
\end{equation}
where $J_a^i = \pm 1$ with equal probability. The solution space
$\mathcal{S}$ of
a large random $K$-SAT formula is non-empty if the constraint density
$\alpha$ is less than a satisfiability threshold $\alpha_s(K)$\cite{Mertens-etal-2006}.
Before $\alpha_s(K)$
is reached, $\mathcal{S}$ has an ergodicity-breaking
transition at a clustering transition point $\alpha_d(K)$, where it
breaks into extremely many solution clusters\cite{Krzakala-etal-PNAS-2007}.
We will see  shortly that this
clustering transition is preceded by another transition at
$\alpha=\alpha_{cm}(K)$, where $\mathcal{S}$ starts
to be heterogeneous as many solution communities are formed.

We use the replica-symmetric cavity method of statistical mechanics\cite{Mezard-Parisi-2001}
to calculate the mean overlap value
$\overline{q}(x)$ at $\alpha < \alpha_d(K)$. As the
partition function Eq.~(\ref{eq:Partition_function}) is a summation
over pairs of solutions $(\vec{\sigma}^{1}, \vec{\sigma}^{2})$,
the state of each vertex is a pair of spins
$(\sigma, \sigma^\prime)$. Consider a vertex $i$ which
is involved in a constraint $a$, $i\in \partial a$. The following two
cavity probabilities $p_{i\rightarrow a}(\sigma_i, \sigma_i^\prime)$ and
$\hat{p}_{a\rightarrow i}(\sigma_i, \sigma_i^\prime)$ are defined: 
$p_{i\rightarrow a}(\sigma_i, \sigma_i^\prime)$ is the
probability that, in the absence of constraint $a$, vertex $i$ has spin
value $\sigma_i$ in solution $\vec{\sigma}^{1}$ and
value $\sigma_i^\prime$ in solution $\vec{\sigma}^{2}$;
and $\hat{p}_{a\rightarrow i}(\sigma_i, \sigma_i^\prime)$ is the probability
that the constraint $a$ is satisfied conditional to
vertex $i$ being in state $(\sigma_i, \sigma_i^\prime)$.
One can write down the
following iterative equations: 
\begin{eqnarray}
\hat{p}_{a\rightarrow i}(\sigma_i, \sigma_i^\prime)
	 &= &
	1- 	\delta_{\sigma_i}^{-J_a^i}
	 \prod\limits_{j\in \partial a \backslash i}
	\Bigl[\sum_{\sigma} p_{j\rightarrow a}(-J_a^j, \sigma)\Bigr]
 -\delta_{\sigma_i^\prime}^{-J_a^i}
	\prod\limits_{j\in \partial a \backslash i}
	\Bigl[\sum_{\sigma} p_{j\rightarrow a}(\sigma, -J_a^j)\Bigr]
	\nonumber \\
& & 		  + \delta_{\sigma_i}^{-J_a^i} \delta_{\sigma_i^\prime}^{-J_a^i}
	\prod\limits_{j\in \partial a\backslash i}
	 p_{j\rightarrow a}(-J_a^j, -J_a^j) \ ,
 	\label{eq:p_a_to_i} \\
p_{i\rightarrow a}(\sigma_i, \sigma_i^\prime)
	&= & C e^{x \sigma_i \sigma_i^\prime}
	\prod\limits_{b\in \partial i\backslash a}
	\hat{p}_{b\rightarrow i}(\sigma_i, \sigma_i^\prime) \ ,
	\label{eq:p_i_to_a}
\end{eqnarray}
where $\delta_m^n$ is the Kronecker symbol,
 $C$ is a normalization constant, and $\partial i$ denotes the
set of constraints that vertex $i$ is associated with. The probability
$p_i(\sigma_i, \sigma_i^\prime)$ of vertex being in the
spin-pair state $(\sigma_i, \sigma_i^\prime)$
has the same expression as
Eq.~(\ref{eq:p_i_to_a}) but with $\partial i\backslash a$ replaced by
$\partial i$. In writing down the above cavity equations, we have
applied the Bethe-Peierls factorization approximation of cavity
probabilities, which corresponds to the replica-symmetric cavity
theory\cite{Mezard-Parisi-2001,Mezard-Montanari-2006}.
For each vertex $i$ the probabilities
$p_i$ and $p_{i\rightarrow a}$ have the symmetry
that $p_i(+,-)=p_i(-,+)$ and 
$p_{i\rightarrow a}(+,-)=p_{i\rightarrow a}(-,+)$.
The mean overlap is 
expressed as
\begin{equation}
 \overline{q}(x)=\frac{1}{N} \sum\limits_{i=1}^{N} 
	\bigl[p_i(+,+)+ p_i(-,-)-2 p_i(+,-)\bigr] \ ,
\end{equation}
and the free entropy density $\phi(x)$  can also be expressed by
the cavity probabilities\cite{Mezard-Parisi-2001}.
The overlap susceptibility $\chi \equiv {\rm d} \overline{q}(x) /{\rm d} x$
is a measures of the overlap fluctuations,
\begin{equation}
	\chi(x) = 
\frac{1}{N}\sum\limits_{i=1}^{N}\sum\limits_{j=1}^{N}
\Bigl[		\langle \sigma_i^{1}\sigma_i^{2} \sigma_j^{1} \sigma_j^{2} \rangle
	- \langle \sigma_i^{1}\sigma_i^{2} \rangle
	\langle \sigma_j^{1} \sigma_j^{2}\rangle\Bigr] \ ,
\end{equation}
where $\langle \ldots \rangle$ means averaging over solution-pairs under
the binding field $x$.

\begin{figure}[t]
	\begin{center}
	\includegraphics[width=0.49\textwidth]{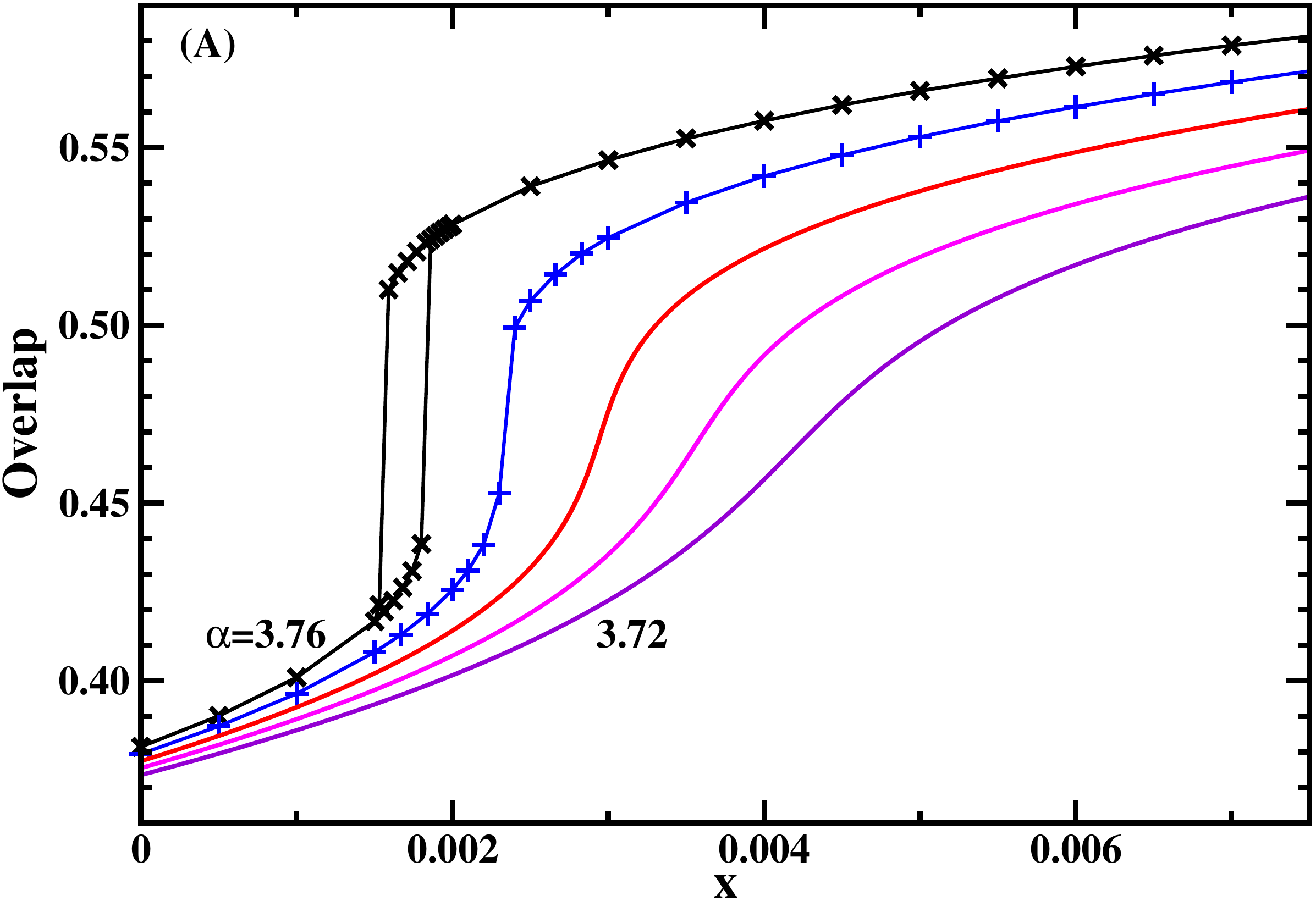}
	\includegraphics[width=0.49\textwidth]{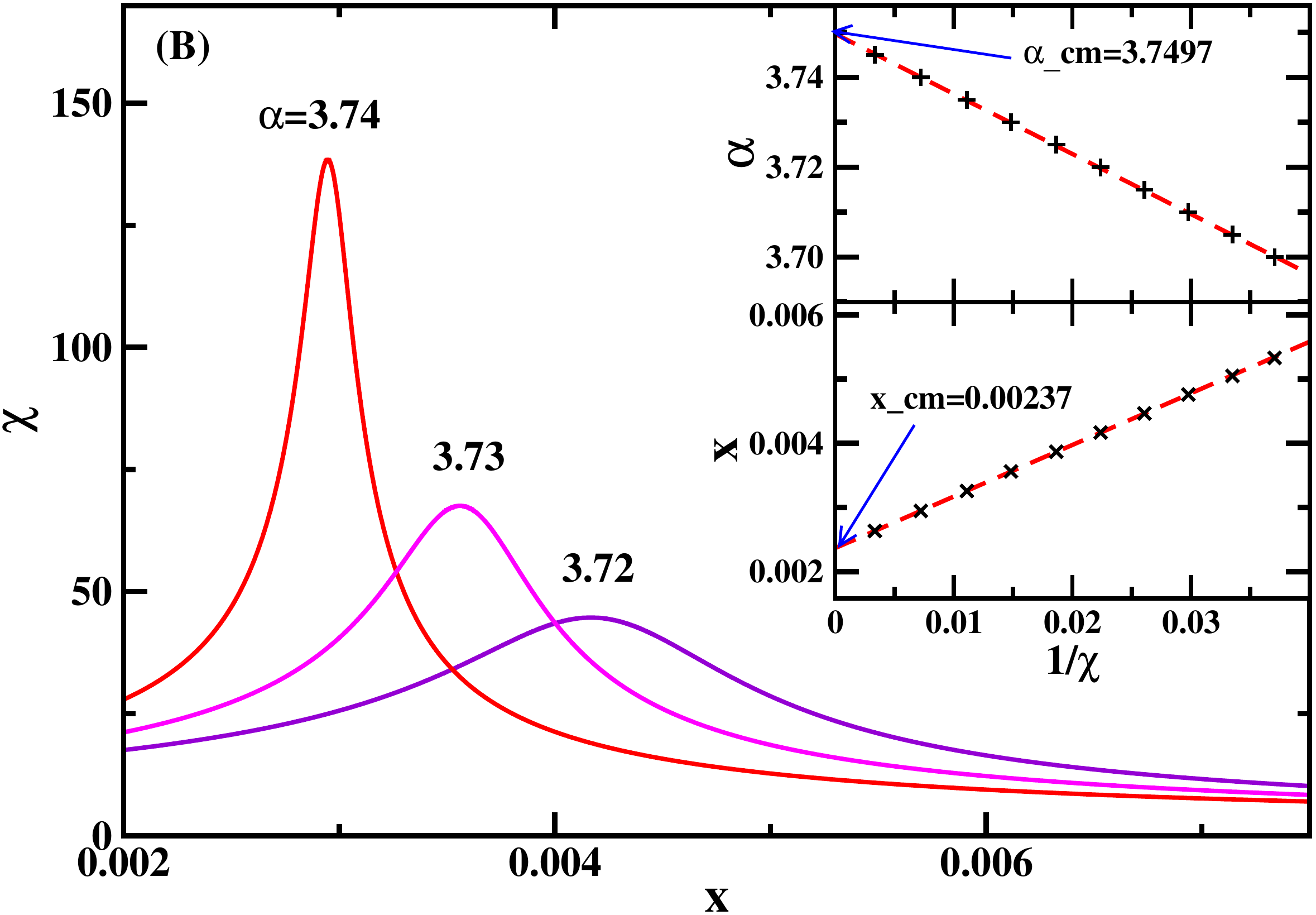}
	\end{center}
	\caption{
	\label{fig:3sat-population}
	The mean overlap $\overline{q}(x)$ (A) and the
	overlap susceptibility $\chi(x)$ (B) at
	different constraint density values $\alpha$ for the random
	$3$-SAT problem. In (A) the value of $\alpha$ increases from
	$3.72$ to $3.76$ with step size $0.01$. 
	The insets of (B) show that the peak value
	of $\chi$ diverges inverse linearly with $\alpha$ and $x$ as
	the critical point $(\alpha_{cm}=3.75, x_{cm}=0.0024)$
	is approached (dashed lines are linear fittings).
	}
\end{figure}

Equations~(\ref{eq:p_a_to_i}) and (\ref{eq:p_i_to_a}) can
be solved by population dynamics method\cite{Mezard-Parisi-2001}.	
Some of the analytical results as obtained for the random $3$-SAT problem are
shown in Fig.~\ref{fig:3sat-population} (the results for $K\geq 4$ are qualitatively
the same). When $\alpha < \alpha_{cm} =3.75$, the mean overlap 
$\overline{q}$ increases  with the binding field $x$  smoothly,
indicating that the solution space $\mathcal{S}$ of the random $3$-SAT problem is
homogeneous. The overlap susceptibility $\chi(x)$ has a single peak,
whose value is inverse proportional to $(\alpha_{cm}-\alpha)$ and
diverges at $\alpha=\alpha_{cm}$ and $x=x_{cm}=0.0024$. The susceptibility $\chi(x)$
is again finite when $\alpha$ exceeds $\alpha_{cm}$, but
 the mean overlap $\overline{q}(x)$
changes discontinuously with $x$ at certain threshold value $x^*$.
This first-order phase transition
at $\alpha > \alpha_{cm}$ suggests that in the
space $\mathcal{S}$ many solution communities (groups of similar
solutions) are formed. For $x> x^*$ the partition function is predominantly
contributed by intra-community solution-pairs (overlap favored), while for $x<x^*$ it is
contributed mainly by inter-community solution-pairs (entropy favored).
The different solution communities of $\mathcal{S}$ all belong to the same
solution cluster ($s(q)$ is non-negative for any $q\in [q_0, 1]$)
as long as $\alpha$ is less than
$\alpha_d=3.87$\cite{Krzakala-etal-PNAS-2007}, but at
$\alpha=\alpha_d$ they
start to break up into different solution clusters ($s(q)$ is not defined
for some intermediate $q$ values\cite{Mora-Mezard-2006}).
 At $\alpha=\alpha_{cm}$
the solution space $\mathcal{S}$ is in a critical state at which the
boundaries between different solution communities disappear.
This situation is qualitatively the same as the critical
state of water at $647$K and  $22.064$MPa, where
the liquid and the gas phase are indistinguishable.

For the random $4$-SAT problem, we find that $\alpha_{cm}=8.4746$, which
is consistent with the simulation results of Ref.\cite{Zhou-Ma-2009}.
The value of $\alpha_{cm}$ is much below the clustering transition point
$\alpha_d=9.38$\cite{Krzakala-etal-PNAS-2007}.

\begin{figure}[t]
	\begin{center}
	\includegraphics[width=0.5\textwidth]{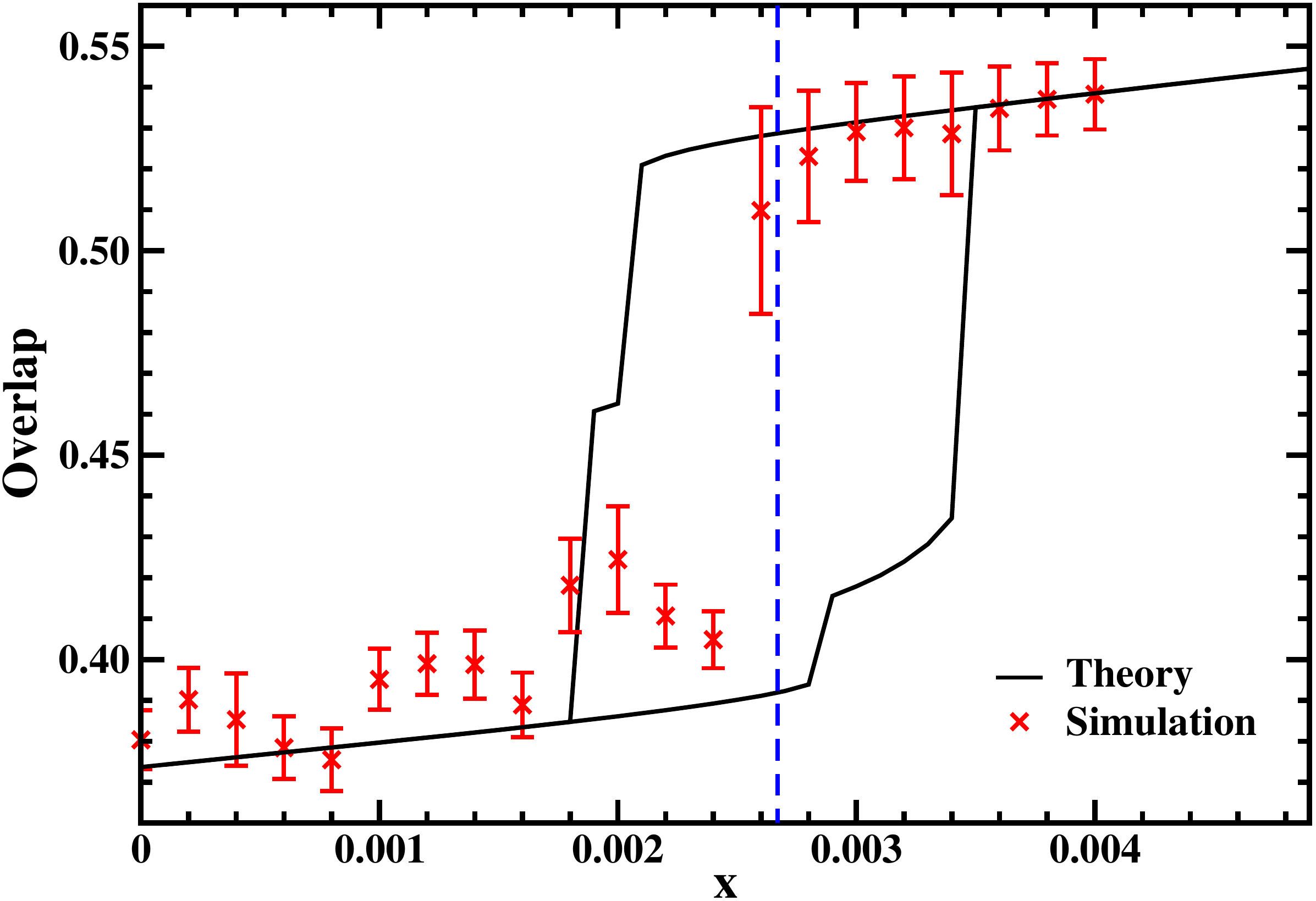}
	\end{center}
	\caption{\label{fig:3sat}	Mean overlap value of solutions with a
	reference solution $\vec{\sigma}^{*}$
	for a single random $3$-SAT formula of $N=10^5$ vertices and constraint density
	$\alpha=3.85$. Solid lines are mean-field analytical result on this single
	formula\cite{Zhou-Ma-2009}, and the symbols with error bars are
	single-spin flips simulation results. Each sampled solution trajectory
	starts from $\vec{\sigma}^*$ and is equilibrated for at least
	$10^7$ Monte Carlo steps
	(each step corresponds to $N$ spin-flip attempts). More than 
	$1000$ overlap values with $\vec{\sigma}^*$ are then sampled at time interval of
	$10^4$ Monte Carlo steps. The blue dashed line marks the equilibrium
	transition value of $x$.}
\end{figure}

The solution space heterogeneity can also be
detected using single solutions as reference points\cite{Zhou-Ma-2009}.
Figure~\ref{fig:3sat} shows the theoretical and
simulation results on a random $3$-SAT formula with $N=10^5$ vertices and
$M=3.85  N$ constraints. The reference solution $\vec{\sigma}^{*}$
is uniformly randomly
sampled from the solution space, and single-spin flips are used in the
simulation to sample solutions $\vec{\sigma}$ with weight proportional to
$\exp[N x q(\vec{\sigma}^{*}, \vec{\sigma})]$\cite{Zhou-Ma-2009}.
The replica-symmetric cavity method predicts an equilibrium
discontinuous change of the mean overlap value with solution
$\vec{\sigma}^{*}$ at $x=x^*=0.00267$, which is confirmed by
simulations. For $x>x^*$ most of the sampled solutions are in the same
solution community of $\vec{\sigma}^{*}$, but for $x< x^*$ the sampled solutions
are scattered in many different solution communities. Because of the high degree
of structural heterogeneity, at $x< x^*$ it takes about $10^7$ Monte Carlo
steps to travel from one solution community to another different community,
making it very difficult to sample independent solutions.

When the solution space of the random $K$-SAT problem becomes heterogeneous
at $\alpha \geq \alpha_{cm}(K)$, the replica-symmetric cavity theory,
which leads to Eqs.~(\ref{eq:p_a_to_i})-(\ref{eq:p_i_to_a}), probably is
not sufficient to describe its statistical properties. In a future publication
we will report the result of the stability analysis on the replica-symmetric
cavity  equations, and present a mean-field study using the first-step
replica-symmetry-breaking cavity theory.


\section{Application to the random $K$-XORSAT problem}

The $K$-XORSAT problem has wide-spread applications
in low-density-parity-check codes\cite{Mezard-Montanari-2009} and is also extremely
studied\cite{RicciTersenghi-Weigt-Zecchina-2001,Mezard-etal-2003,Cocco-etal-2003,Mora-Mezard-2006}.
The constraint energy $E_a$ of this model is expressed in
Eq.~(\ref{eq:xorsat}), where $J_a=\pm 1$ with equal probability. The
solution space of a random $K$-XORSAT problem breaks into exponential solution
clusters of equal size at a clustering transition point
$\alpha= \alpha_d(K)$\cite{Mezard-etal-2003,Cocco-etal-2003}. We have applied the
replica-symmetric cavity method to this problem and obtained the same qualitative
results as for the random $K$-SAT problem, namely that before the ergodicity
of the solution is broken, exponentially many solution communities start to
form in the solution space as the constraint density reaches a critical value
$\alpha_{cm}(K)$. At $\alpha=\alpha_{cm}(K)$ the solution space is in a
critical state. For $K=3$ we find that $\alpha_{cm}(3)=0.6182$, which is much lower than
the value of $\alpha_d(3)=0.818$\cite{RicciTersenghi-Weigt-Zecchina-2001}. For the
random $4$-XORSAT problem, we find that $\alpha_{cm}(4)=0.504$,
while $\alpha_{d}(4)=0.772$\cite{Mezard-etal-2003}.

The random $K$-XORSAT problem has a gauge symmetry that can be exploited to
simply the mean-field calculations\cite{Mora-Mezard-2006}.
Suppose $\vec{\sigma}^{1}$ is a solution, we can perform a gauge transformation
$\sigma_i\rightarrow \tilde{\sigma}_i = \sigma_i \sigma_i^{1}$ to change the
constraint energy Eq.~(\ref{eq:xorsat}) into $E_a = 1- \prod_{i\in \partial a} 
\tilde{\sigma}_i$. All the coupling constants $J_a$ then become unity. The
solution space structure of the random $K$-SAT problem looks the same from
any a reference solution. We have used this nice property to calculate the
total number of solutions that have a overlap value $q$ with a randomly
chosen reference solution.


\section{Discussion}

The main conclusion of this work is that, the solution space of a
random constraint satisfaction problem has a transition to structural heterogeneity
at a critical constraint density $\alpha_{cm}$, where many solution communities
form. These solution communities serve as precursors for the splitting of
the solution space into many  solution clusters at a larger threshold value
$\alpha_{d}$ of constraint density. This work brings a refined picture on how
ergodicity of the solution space of a CSP  finally breaks as the constraint
density increases.

\begin{figure}[t]
	\begin{center}
	\includegraphics[width=0.4\textwidth]{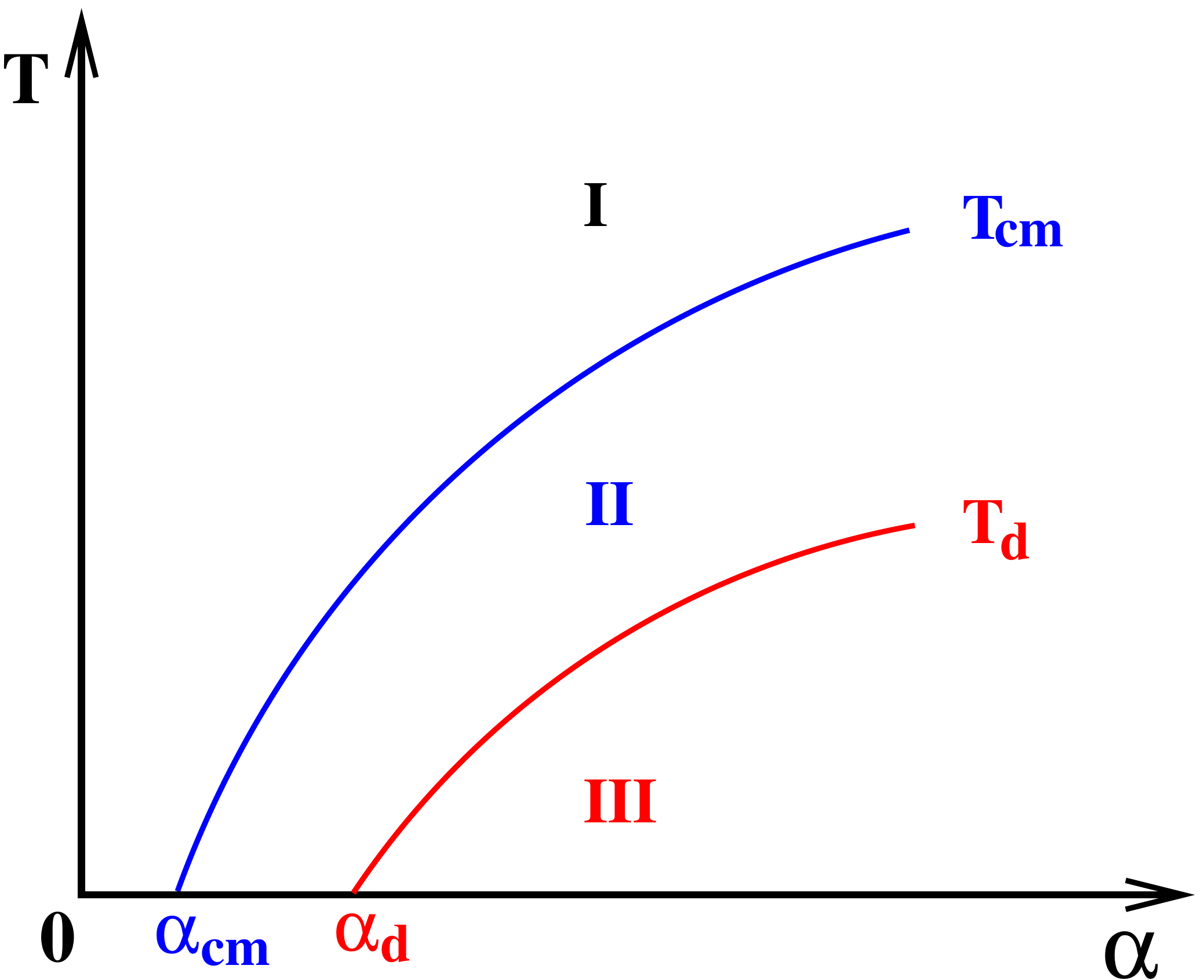}
	\end{center}
	\caption{\label{fig:phase}
	Schematic phase diagram for a constraint satisfaction problem, using
	temperature $T$ and constraint density $\alpha$ as control parameters.
	The configuration space is homogeneous and ergodic in region I.
	As the temperature $T$ decreases to $T_{cm}(\alpha)$,
	a homogeneity-breaking transition occurs, and the
	configuration space becomes non-homogeneous but still ergodic (region II).
	As $T$ further decreases to $T_{d}(\alpha)$,
	an ergodicity-breaking (clustering) transition occurs, and the
	configuration space breaks into many separated clusters (region III). At $T=0$,
	the ground-state configuration space  is non-homogeneous
	at $\alpha \geq \alpha_{cm}$ and  non-ergodic at $\alpha\geq \alpha_d$.
}
\end{figure}

In spin-glass models with multi-spin interactions, the control parameter is often the
temperature. The method presented here can also be used to study how the configuration
spaces of these systems evolve with temperature. We suggest  that similar
heterogeneity transitions will occur before the clustering (or dynamical) transition.
The following scenario is expected (see Fig.~\ref{fig:phase}): at high temperatures the
configuration space of a spin-glass or a lattice glass model system is
in a homogeneous phase; as the temperature $T$ decreases to certain
critical value $T_{cm}$, many communities of configurations form in the
configuration space, and the configuration space is then in a heterogeneous but
still ergodic phase; as $T$ decreases further to $T_d$, the different configuration
communities separate into different Gibbs states, and the configuration space is
no longer ergodic. The values of $T_{cm}$ for the random $K$-SAT problem and the 
random $K$-XORSAT problem as a function of the constraint density $\alpha$ will be
calculated in a forthcoming publication. A related study was reported by
Krzakala and Zdeborova recently on the
the adiabatic evolution of single
Gibbs states of a spin-glass system as a function of temperature\cite{Krzakala-Zdeborova-2009,Zdeborova-Krzakala-2010}.

As the solution space of a CSP or the configuration space of a spin-glass or
lattice glass system becomes heterogeneous and the configurations aggregate into many
different communities, a stochastic search process based only on local rules (e.g.,
solution space random walking\cite{Zhou-2009}) or a local dynamical process
(e.g., single-particle heat-bath dynamics of a lattice glass\cite{Darst-Reichman-Biroli-2009})
may get slowing down considerably and
show heterogeneous behavior. The configuration space heterogeneity discussed
in this paper probably is deeply connected to the phenomenon of spatial
dynamical heterogeneity of glass-forming liquids\cite{Ediger-2000,Cavagna-2009}.
This research direction will be pursued in future work.


\section*{Acknowledgement}

HZ thanks Hui Ma and Ying Zeng for discussions and Lenka Zdeborova for
help comments on an earlier version of the manuscript.
This work was partially supported by the
National Science Foundation of China
(Grant number 10774150) and the China 973-Program (Grant number 2007CB935903).


\end{document}